\documentclass[10pt,aps,prl,onecolumn,notitlepage,superscriptaddress,preprintnumbers,longbibliography]{revtex4-1}

\usepackage{amsmath,amssymb,amsfonts}
\usepackage{bm}
\usepackage{graphicx,color}
\usepackage{verbatim}
\usepackage{float}
\usepackage{dcolumn}
\usepackage{natbib}
\usepackage{hyperref}
\usepackage{mdframed}
\definecolor{ly}{RGB}{255,255,200}

\newcommand{\ie}{\emph{i.e.}}
\newcommand{\eg}{\emph{e.g.}}

\newcommand{\ER}{Erd\"{o}s-R\'enyi }
\newcommand{\avg}[1]{\langle #1\rangle}
\newcommand{\vc}[1]{\bm{#1}}
\newcommand{\mx}[1]{\bm{#1}}

\begin{document}

\title{The Statistical Physics of Real-World Networks}

\author{Giulio Cimini}
\affiliation{IMT School for Advanced Studies, Piazza San Francesco 19, 55100 Lucca (Italy)}
\affiliation{Istituto dei Sistemi Complessi (CNR) UoS Sapienza, Dipartimento di Fisica, ``Sapienza'' Universit\`a di Roma, P.le A. Moro 2, 00185 Rome (Italy)}
\author{Tiziano Squartini}
\affiliation{IMT School for Advanced Studies, Piazza San Francesco 19, 55100 Lucca (Italy)}
\author{Fabio Saracco}
\affiliation{IMT School for Advanced Studies, Piazza San Francesco 19, 55100 Lucca (Italy)}
\author{Diego Garlaschelli}
\affiliation{IMT School for Advanced Studies, Piazza San Francesco 19, 55100 Lucca (Italy)}
\affiliation{Lorentz Institute for Theoretical Physics, Leiden University, Niels Bohrweg 2, 2333 CA Leiden (The Netherlands)}
\author{Andrea Gabrielli}
\affiliation{Istituto dei Sistemi Complessi (CNR) UoS Sapienza, Dipartimento di Fisica, ``Sapienza'' Universit\`a di Roma, P.le A. Moro 2, 00185 Rome (Italy)}
\affiliation{IMT School for Advanced Studies, Piazza San Francesco 19, 55100 Lucca (Italy)}
\author{Guido Caldarelli}
\affiliation{IMT School for Advanced Studies, Piazza San Francesco 19, 55100 Lucca (Italy)}
\affiliation{Istituto dei Sistemi Complessi (CNR) UoS Sapienza, Dipartimento di Fisica, ``Sapienza'' Universit\`a di Roma, P.le A. Moro 2, 00185 Rome (Italy)}
\affiliation{European Centre for Living Technology, Universit\`a di Venezia ``Ca' Foscari'', S. Marco 2940, 30124 Venice (Italy)}

\begin{abstract}
In the last 15 years, statistical physics has been a very successful framework to model complex networks. On the theoretical side, this approach has brought novel insights into a variety of physical phenomena, 
such as self-organisation, scale invariance, emergence of mixed distributions and ensemble non-equivalence, that display unconventional features on heterogeneous networks. 
At the same time, thanks to their deep connection with information theory, statistical physics and the principle of maximum entropy 
have led to the definition of null models for networks reproducing some features of real-world systems, but otherwise as random as possible. 
We review here the statistical physics approach and the various null models for complex networks, focusing in particular on the analytic frameworks reproducing the local network features. 
We then show how these models have been used to detect statistically significant and predictive structural patterns in real-world networks, 
as well as to reconstruct the network structure in case of incomplete information. 
We further survey the statistical physics models that reproduce more complex, semi-local network features using Markov chain Monte Carlo sampling, 
as well as the models of generalised network structures such as multiplex networks, interacting networks and simplicial complexes.
\end{abstract}

\maketitle

The science of networks has exploded in the Information Age thanks to the unprecedented production and storage of data on basically any human activity. 
Indeed, a network represents the simplest yet extremely effective way to model a large class of technological, social, economic and biological systems: 
a set of entities (nodes) and of interactions (links) among them. These interactions do represent the fundamental degrees of freedom of the network, 
and can be of different types---undirected or directed, binary or valued (weighted)---depending on the nature of the system and the resolution used to describe it. 
Notably, most of the networks observed in the real world fall within the domain of complex systems, as they exhibit strong and complicated interaction patterns, 
and feature collective emergent phenomena that do not follow trivially from the behaviours of the individual entities \cite{dorogovtsev2008critical}. 
For instance, many networks are {\em scale-free} \cite{barabasi1999emergence}, meaning that the number of links incident to a node (known as the node's {\em degree}) follows a power-law distribution: 
most of the nodes have a few links, but a few of them (the hubs) are highly connected. 
The same happens for the distribution of the total weight of connections incident to a node (the node's {\em strength}) \cite{yook2001weighted,barrat2004weighted}. 
In addition, most real-world networks are organised into modules or display a community structure \cite{newman2004finding,fortunato2010community}, 
and they possess high clustering---as nodes tend to create tightly linked groups, but are also {\em small-world} \cite{watts1998collective,amaral2000classes,chung2002average} 
as the mean distance (in terms of number of connections) amongst node pairs scales logarithmically with the system size. 
The observation of these universal features in complex networks has stimulated the development of a unifying mathematical language 
to model their structure and understand the dynamical processes taking place on them---such as the flow of traffic on the Internet 
or the spreading of either diseases or information in a population \cite{albert2002statistical,newmann2003structure,boccaletti2006complex}. 

Two different approaches to network modelling can be pursued. The first one consists in identifying one or more microscopic mechanisms driving the formation of the network, 
and use them to define a dynamic model which can reproduce some of the emergent properties of real systems. The small-world model \cite{watts1998collective}, 
the preferential attachment model \cite{barabasi1999emergence}, the fitness model \cite{bianconi2001bose,caldarelli2002scalefree,dorogovtsev20004structure}, 
the relevance model \cite{medo2011temporal} and many others follow this approach 
which is akin to kinetic theory. These models can handle only simple microscopic dynamics, and thus while providing good physical insights they need several refinements to give quantitatively accurate predictions.

The other possible approach consists in identifying a set of characteristic static properties of real systems, and then building networks having the same properties 
but otherwise maximally random. This approach is thus akin to statistical mechanics and therefore is based on rigorous probabilistic arguments 
that can lead to accurate and reliable predictions. The mathematical framework is that of {\em exponential random graphs} (ERG), which has been first introduced in the social sciences and statistics 
\cite{holland1981exponential,frank1986markov,strauss1986general,wasserman1996logit,anderson1999primer,snijders2006new,robins2007introduction,cranmer2011inferential,snijders2011statistical} 
as a convenient formulation relying on numerical techniques such as Markov chain Monte Carlo algorithms. The interpretation of ERG in physical terms is due to Park and Newman \cite{park2004statistical}, 
who showed how to derive them from the principle of maximum entropy and the statistical mechanics of Boltzmann and Gibbs. 

As formulated by Jaynes \cite{jaynes1957information}, the variational principle of maximum entropy states that the probability distribution best representing the current state of (knowledge on) a system 
is the one which maximises the Shannon entropy, subject in principle to any prior information on the system itself. This means making self-consistent inference assuming maximal ignorance 
about the unknown degrees of freedom of the system \cite{shore1980axiomatic}. The maximum entropy principle is conceptually very powerful and finds almost countless applications 
in physics and in science in general \cite{presse2013principles}. In the context of network theory, the ensemble of random graphs with given aggregated (macroscopic or mesoscopic) 
structural properties derived from the maximum entropy approach has a two-sided important application. On one hand, when the actual microscopic configuration of a real network is not accessible, 
this ensemble describes the most probable network configuration: as in traditional statistical mechanics, the maximum entropy principle allows to gain maximally unbiased information 
in the absence of complete knowledge. On the other hand, when the actual microscopic configuration of the network is known, this ensemble defines a null model which allows to assess the significance 
of empirical patterns found in the network---against the hypothesis that the network structure is determined solely by its aggregated structural properties. 

The purpose of this review is to present the theoretical developments and empirical applications for the statistical physics of real-world complex networks. 
We start by introducing the general mathematical formalism, and then we focus on the analytic models obtained by imposing mesoscopic (\ie, local) constraints, 
highlighting the novel physical concepts that can be learned from such models. After that we present the two main fields of applications for these models: 
the detection of statistically significant patterns in empirical networks, and the reconstruction of network structures from partial information. 
At the end we discuss the models obtained by imposing semi-local network features, as well as the most recent developments on generalised network structures and simplices.

\subsection*{Statistical mechanics of networks}

\begin{figure*}[h]
\includegraphics[width=\textwidth]{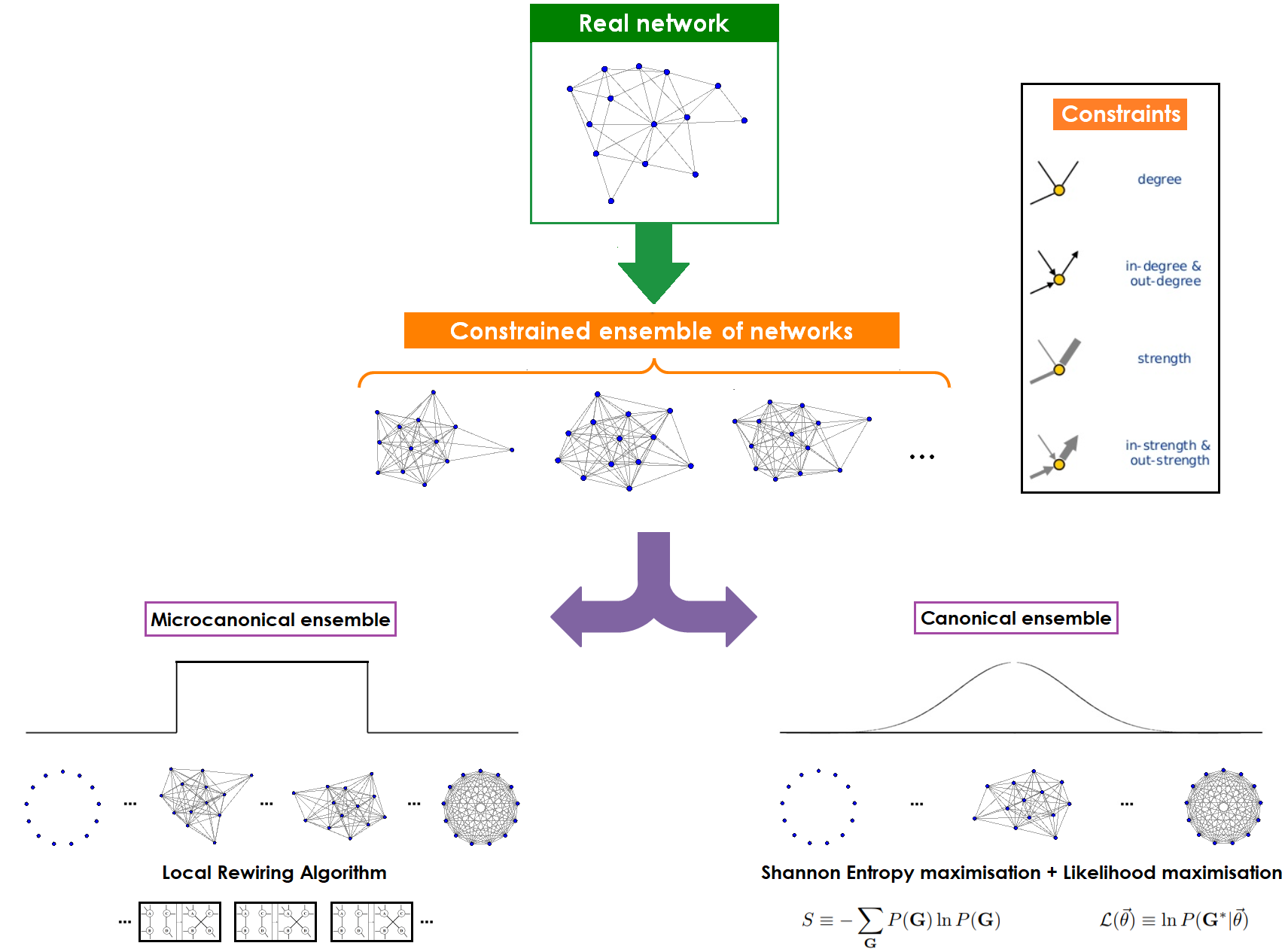}
\caption{Construction of the microcanonical and canonical ensemble of networks from local constraints---in this case, the degrees $\vc{k}^*$ of a real network $\mx{G}^*$. 
On the left, the microcanonical approach relies on the link rewiring method to numerically generate several network configurations, each with exactly the same degree sequence of $\mx{G}^*$: 
$P(\mx{G})$ is non-zero (and uniform, provided the sampling is unbiased) only for the subset of graphs that realise the enforced constraints exactly. 
On the right, the canonical approach obtains $P(\mx{G})$ by maximising the Shannon entropy constraining the expected degree values within the ensemble, 
and then maximising the Likelihood of $P(\mx{G}^*)$ to find the ensemble parameters $\vc{\theta}^*$ such that the expected degree values match the observations in $\mx{G}^*$. 
Thus $P(\mx{G})$ is non-zero for any graph (ranging from the empty to the complete one). Figure adapted from \cite{squartini2011analytical}.}
\label{fig:maxE}
\end{figure*}

The statistical physics approach defined by ERG consists in modelling a network system $\mx{G}^*$ 
through an ensemble $\Omega$ of graphs with the same number $N$ of nodes and type of links of $\mx{G}^*$ (Fig.~\ref{fig:maxE}). 
The model is specified by $P(\mx{G})$, the occurrence probability of a graph $\mx{G}\in\Omega$. 
According to statistical mechanics and information theory \cite{jaynes1957information,jaynes1982rationale}, in order to achieve the most unbiased expectation 
about the microscopic configuration of the system under study, such probability distribution is the one maximising the Shannon entropy
\begin{equation}
\mathcal{S}=-\sum_{\mx{G}\in\Omega}P(\mx{G})\ln P(\mx{G})\label{eq:entropy},
\end{equation}
subject to the normalisation condition $\sum_{\mx{G}\in\Omega}P(\mx{G})=1$, as well as to a collection of constraints $\vc{c}^*$ 
representing the macroscopic properties enforced on the system (and thus defining the sufficient statistics of the problem). 

Imposing hard constraints, that is, assigning uniform $P(\mx{G})$ over the set of graphs that satisfy $\vc{c}(\mx{G})=\vc{c}^*$ and zero probability to graphs that do not leads to the {\em microcanonical ensemble}. 
Typically, this ensemble is resilient to analytic treatment beyond steepest descent approximations \cite{bianconi2008entropy}, and it is thus sampled numerically (see Box~1). 

The {\em canonical ensemble} is instead obtained by imposing soft constraints, that is, by fixing the expected values of the constraints over the ensemble, $\sum_{\mx{G}\in\Omega}\vc{c}(\mx{G})P(\mx{G})=\vc{c}^*$. 
Introducing the set of related Lagrange multipliers $\vc{\theta}$, the constrained entropy maximisation returns 
\begin{equation}
P(\mx{G}|\vc{\theta})=e^{-H(\mx{G},\vc{\theta})}/Z(\vc{\theta})\label{eq:prob}
\end{equation}
where $H(\mx{G},\vc{\theta})=\vc{\theta}\cdot\vc{c}(\mx{G})$ is the Hamiltonian and $Z(\vc{\theta})=\sum_{\mx{G}\in\Omega}e^{-H(\mx{G},\vc{\theta})}$ is the partition function. 
Thus the canonical $P(\mx{G}|\vc{\theta})$ depends on $\mx{G}$ only through $\vc{c}(\mx{G})$, which automatically implies that graphs with the same value of the constraints have equal probability. 
This means that the canonical ensemble is maximally non-committal with respect to the properties that are not enforced on the system \cite{squartini2015unbiased}. 

Remarkably, the microcanonical and canonical ensembles turn out to be non-equivalent in the thermodynamic limit $N\to\infty$ for models of networks with an extensive number of constraints 
\cite{anand2009entropy,squartini2015breaking,squartini2018reconnecting}. This is in contrast to what happens in traditional statistical physics (except possibly at phase transitions), 
where the number of constraints---\eg, total energy and total number of particles---is typically finite. In this sense, complex networks not only provide an important application domain for statistical physics, 
but can even expand our fundamental understanding of statistical physics itself, leading to novel theoretical insight. Additionally, from a practical point of view, 
the breaking of ensemble equivalence in networks implies that the choice between microcanonical and canonical ensembles cannot be based solely on mathematical convenience as usually done, 
and rather should follow from a principled criterion. In particular, since the canonical ensemble describes systems subject to statistical fluctuations, its use is more appropriate 
when the observed values of the constraints can be affected by measurement errors, missing and spurious data, or simply stochastic noise. 
Luckily, as we shall see, this criterion leads to the more analytically tractable ensemble.

The definition of the canonical ensemble of eq. \eqref{eq:prob} specifies the functional form of $P(\mx{G}|\vc{\theta})$, 
but leaves the Lagrange multipliers as parameters to be determined by the constraints equations $\sum_{\mx{G}\in\Omega}\vc{c}(\mx{G})P(\mx{G}|\vc{\theta})=\vc{c}^*$. 
Since in practical applications the average values of the constraints are seldom available, a possible strategy is to draw the Lagrange multipliers from chosen probability densities 
inducing archetypal classes of networks (\eg, regular graphs, scale-free networks, and so on) \cite{park2004statistical,garlaschelli2009generalized,bianconi2008entropy}. 
When instead the task is to fit the model to the observations $\vc{c}^*\equiv\vc{c}(\mx{G}^*)$ for a {\em given} empirical network $\mx{G}^*$, 
the optimal choice to find the values $\vc{\theta}^*$ is to maximise the likelihood functional \cite{garlaschelli2008maximum,squartini2011analytical}
\begin{equation}
\mathcal{L}(\vc{\theta})=\ln P(\mx{G}^*|\vc{\theta}).\label{eq:likelihood}
\end{equation}
This results in the matching $\sum_{\mx{G}\in\Omega}\vc{c}(\mx{G})P(\mx{G}|\vc{\theta}^*)\equiv\vc{c}(\mx{G}^*)$ between the ensemble average and the observed value of each constraint.

\begin{mdframed}[backgroundcolor=ly]
\subsection*{Box 1: alternative ensemble constructions}
Various methods to define ensembles of graphs with local constraints, alternative to maximum entropy, have been proposed in the literature. 
Here we briefly present them for the case of binary undirected graphs, which is by far the simplest and most frequently explored situation. 

Computational methods explicitly generate several random networks with the desired degree sequence. 
The `bottom-up' approach initially assigns to each node a number of link stubs equal its target degree; 
then, pairs of stubs are randomly matched avoiding the formation of self-loops and multi-links \cite{newman2001random,itzkovitz2004reply,catanzaro2005generation}. 
Unfortunately, very often this procedure gets stuck in configurations where nodes requiring additional connections have no more eligible partners, 
leading to unacceptably many sample rejections \cite{maslov2004detection}.
The `top-down' approach instead starts from a realised network and generates a set of randomised variants by iteratively applying a \emph{link rewiring algorithm} 
that preserves the degree distribution \cite{maslov2002specificity,milo2002network,zamora2008reciprocity,zlatic2009rich-club}. 
The drawbacks here is that the numbers of rewirings needed to generate a single configuration is very large and not rigorously specified \cite{tabourier2011generating}. 
Additionally, the algorithm may fail to sample the ensemble uniformly, unless employing a rewiring acceptance probability 
depending on the current network configuration \cite{artzy2005generating,coolen2009constrained,roberts2012unbiased,carstens2017switching}. 
Other methods rely on theorems setting necessary and sufficient conditions for a degree sequence to be \emph{graphic}, \ie, realised by at least one graph, 
and exploit such conditions to define biased sampling algorithms and sampling probabilities \cite{delgenio2010efficient,blitzstein2011sequential,kim2012constructing}. 
These approaches are however rather costly, especially for highly heterogeneous networks. 

Analytic methods instead define (usually approximated) explicit expressions for expected values of network properties as a function of the imposed constraints. 
A standard approach relies on the {\em generating function} $g(z)=\sum_k z^k P(k)$ for the degree distribution \cite{newman2001random,newman2009random}, 
yet is rigorously defined only for infinite and locally tree-like networks---even if it often works surprisingly well for real networks \cite{melnik2011unreasonable}. 
An alternative popular approach is based on the explicit expression $p_{ij}=\tfrac{1}{2}k_i^*k_j^*/E^*$ for the connection probability between any two nodes $i$ and $j$ in the randomised ensemble 
($E^*$ is the total number of links) \cite{chung2002connected}. This model actually defines systems with self-loops as well as multilinks \cite{park2003origin,serrano2005weighted,garlaschelli2008maximum}. 
In particular, since $p_{ij}$ may exceed $1$ for pairs of high-degree nodes, the model requires that $i$ and $j$ should be connected by more than one link to actually realise the imposed constraints. 
The occurrence of these events in not negligible in scale-free networks with $P(k)\sim k^{-\gamma}$ 
for which the natural cutoff $\sim N^{1/(\gamma-1)}$ is larger than the structural cut-off ($\sim N^{1/2}$ for uncorrelated networks) \cite{burda2003uncorrelated,boguna2004cuttoffs}.
\medskip
\end{mdframed}

\paragraph*{\bf \em Imposing local constraints} --- Unlike most alternative approaches (briefly outlined in Box 1), the maximum entropy method is general and works for networks that are either 
binary or weighted, undirected or directed, sparse or dense, tree-like or clustered, small or large. However, the specification of the occurrence probability of a graph in the ensemble can be a challenging task. 
The point is that, like in conventional equilibrium statistical mechanics, whether the partition function can be analytically computed depends on the particular constraints imposed. 
In a handful of lucky cases such computation is indeed feasible, so that expectation values and higher moments of any quantity in the ensemble can be analytically derived. 
Besides very simple models like the well-known \ER random graph \cite{erdos1959random}, this happens for the important constraints describing the {\em local} network structure 
from the viewpoint of each individual node---namely, degrees~$\vc{k}^*$ and strengths~$\vc{s}^*$ \cite{park2004statistical}. The scale-free behaviour observed for these quantities in real-world systems 
is indeed the most elemental signature distinguishing networks from systems typically studied in physics such as gases, liquids, lattices, 
and cannot be obtained from simple models with global constraints. For instance, the \ER model is obtained in the ERG formalism by constraining the expected total number of links. 
This leads to an ensemble in which each pair of nodes is connected with fixed probability $p$, so that the degree distribution follows a binomial law. 
Thus in order to construct ERG ensembles that are both practically useful and theoretically sound (by accurately replicating the observed heterogeneity of real-world networks), 
imposing local constraints separately for each node is a minimum requirement. And local constraints make the method analytic because of the {\em independence of dyads}: 
$P(\mx{G})$ factorises into link-specific terms, whose contribution to the partition function sum can be evaluated independently from the rest of the network.
Note that local constraints lay at the {\em mesoscopic} level between the microscopic degrees of freedom of the network (the individual links) and the macroscopic aggregation of all degrees of freedom 
into global quantities like the total number of links---corresponding for instance to the total energy of the system in traditional statistical physics. 

The ERG model obtained by constraining the degrees $\vc{c}^*\equiv\vc{k}^*$ is known as the (canonical) \emph{binary configuration model} (BCM). 
In the simplest undirected case, the entropy maximisation procedure returns an ensemble connection probability between any two nodes $i$ and $j$ given by 
\begin{equation}
p_{ij}=\frac{x_ix_j}{1+x_ix_j},\label{eq:cm}
\end{equation}
where $\vc{x}$ are the (exponentiated) Lagrange multipliers \cite{park2004statistical}. 
The {\em weighted configuration model} (WCM) \cite{serrano2005weighted} is instead obtained by constraining the strengths $\vc{c}^*\equiv\vc{s}^*$. 
Again in the simpler undirected case, and considering integer weights, the connection probability between any two nodes $i$ and $j$ is given by $p_{ij}=y_iy_j$, 
where $\vc{y}$ are the (exponentiated) Lagrange multipliers. 
The probability distribution and the ensemble average for the weight of that link (or, equivalently, for how many links are established between the two nodes) are 
$q_{ij}(w)=(y_iy_j)^w(1-y_iy_j)$ and 
\begin{equation}
\avg{w_{ij}}=\frac{y_iy_j}{1-y_iy_j}.\label{eq:wcm}
\end{equation}
These models naturally recall the traditional statistical mechanics for systems of non-interacting particles, 
once connections are interpreted as particles in a quantum gas and pairs of nodes as single-particle states. 
Indeed in binary networks each single-particle state can be occupied by at most one particle, so the resulting statistics of eq. \eqref{eq:cm} is fermionic, 
whereas, in weighted networks single-particle states can be occupied by an arbitrary number of particles, so that eq. \eqref{eq:wcm} better describes a system of bosons---where 
Bose-Einstein condensation can occur between very strong nodes for which $y_iy_j\to1$ \cite{park2004statistical}. Notably, a mixed Bose-Fermi statistics is obtained 
when degrees and strengths are imposed simultaneously \cite{garlaschelli2009generalized}, as in the {\em enhanced configuration model} (ECM) \cite{mastrandrea2014enhanced}. 
Again in the simplest undirected case and using (exponentiated) Lagrange multipliers $\vc{x}$ and $\vc{y}$ respectively for degrees and strengths, 
in this case one gets $p_{ij}=(x_ix_jy_iy_j)/(x_ix_jy_iy_j-y_iy_j+1)$ and $q_{ij}(w>0)=p_{ij}(y_iy_j)^{w-1}(1-y_iy_j)$. 
Hence the ECM differs from the WCM in the way the first link established between any two nodes is treated: 
the processes of creating a connection from scratch and that of reinforcing an existing one obey intrinsically different rules, 
the former meant to satisfy the degree constraints and the second to fix the values of the strengths. 
Like the ensemble non-equivalence, this mechanism and the resulting mixed statistics 
constitutes a novel physical phenomenon that the statistical physics approach to networks can unveil.

\subsection*{Patterns validation}

Validating models, that is comparing their statistical properties with measurements of real-world systems, is an essential step in the activity of theoretical physicists. 
Specifically, in the context of networks and complex systems, besides looking for what a model is able to explain, 
much research has been devoted to identify the empirical properties which deviate from a benchmark model 
\cite{maslov2002specificity,park2003origin,barrat2004architecture,maslov2004detection,colizza2006rich,serrano2006correrlations,guimera2006classes,bhattacharya2008international,opsahl2008prominence}. 
This is because possible deviations likely bear important information about the unknown formation process or a particular function of the empirical network. 

Maximum entropy models are perfectly suited for this task. Starting from a real network $\mx{G}^*$, they derive the null hypothesis (that is, the benchmark model) from the set of 
properties $\vc{c}(\mx{G}^*)$ imposed as constraints, and otherwise assuming no other information on the system. 
This means formulating the null hypothesis that these constraints are the only explanatory variables for the network at hand. 
The other properties of $\mx{G}^*$ can then be statistically tested and possibly validated against this null hypothesis. 
For instance, a null model derived from imposing the total number of links as (macroscopic) constraint is typically used to reject a homogeneity hypothesis for the degree distribution. 
Instead, when imposing local constraints the aim is to check whether higher-order patterns of a real network---such as reciprocity 
(the tendency of nodes in a directed network to be mutually linked), clustering (the tendency of node triples to be connected together forming triangles), 
(dis-)assortativity (the tendency of nodes to be linked to other nodes with (dis-)similar degrees)---are 
statistically significant beyond what can be expected from the heterogeneity of degrees or strengths themselves. 
Note that the possibility to analytically characterise the ensemble given by the use of local constraints means that expectation values and standard deviations of most quantities of interests 
can be explicitly derived, so that hypothesis testing based on standard scores can be easily performed \cite{squartini2011analytical}. And when the ensemble distribution of the considered quantity is not normal, 
sampling of the configuration space using the explicit formulas for $P(\mx{G})$ easily allows to perform statistical tests based on $p$-values. 

In the context of patterns validation, one of the most studied systems---that we also discuss here as an illustrative example---has been the World Trade Web (WTW), 
namely the network of trade relationships between countries in the world \cite{serrano2003topology,garlaschelli2004fitness}. The observed disassortative pattern, 
for which countries with many trade partners are connected on average to countries with few partners \cite{garlaschelli2005structure,fagiolo2009world}, turns out to be statistically explained 
with good approximation by the degree sequence \cite{squartini2011analytical}. The same happens with the clustering coefficient \cite{squartini2011analytical}. 
These two observations exclude the presence of meaningful indirect (economic) interactions on top of a concatenation of direct (economic) interactions. 
Things change however when the weighted version of the network is analysed---although no unique extension of binary quantities exists in this case 
\cite{barrat2004architecture,newman2004analysis,ahnert2007ensemble}. Indeed, the disassortative pattern is still observed, but is not compatible with that of the WCM null model. 
Concerning the weighted clustering coefficient \cite{saramaki2007generalization}, the agreement between empirical network and model is only partial. 
These findings point to the fact that, unlike in the binary case, the knowledge of the strengths conveys only limited information about the higher-order weighted structure of the network. 
This, together with the fact that also basic topological properties like the link density (\ie, the fraction of possible connections that are actually realised in the network) 
are not reproduced by the WCM, suggests that even in weighted analyses the binary structure plays an important role, irreducible to what local weighted properties can explain. 

\paragraph*{\bf \em Network motifs and communities} --- Motifs \cite{milo2002network,shenorr2002network} are patterns of interconnections involving a small subset of nodes in the network, 
thus generalising the clustering coefficient. The typical null model used to study motifs in directed networks is obtained by constraining, in addition to degrees, also the number of reciprocal links per node 
\cite{garlaschelli2004patterns,garlaschelli2006multispecies,squartini2012triadic}, which is meaningful in many contexts. 
For instance, in food webs the presence of bi-directed predator-prey relations between two species strongly characterises an ecosystem \cite{stouffer2007evidence}. 
In interbank networks, the presence of mutual loans between two banks is a signature of trust between them, and in fact 
the appearance of the motif corresponding to three banks involved in a circular lending loop with no reciprocation provides significant early warnings of financial turmoil \cite{squartini2013early}. 

Community structure instead refers to the presence of large groups of nodes that are densely connected internally but sparsely connected externally \cite{fortunato2010community}. 
Most of the methods to find communities in networks are based on the optimisation of a functional, the {\em modularity} \cite{newman2004finding} being the most prominent example, 
that compares the number of links falling within and between groups with the expectation of such numbers under a given null network model. 
This comparison with the null model is a fundamental step of the procedure, since even random graphs possess an intrinsic yet trivial community structure 
\cite{guimera2004modularity,recihardt2007partitioning}. Indeed, in its original formulation the modularity is defined on top of the configuration model by Chung and Lu 
\cite{chung2002connected} (see Box 1). This model is fast and analytic but generates self-loops and multilinks, hence it gives an accurate benchmark only when these events are rare 
(\ie, in very large and sparse networks). More generally the maximum entropy approach (\eg, the configuration model of eq. \eqref{eq:cm}), while more demanding from a practical viewpoint, 
can provide the proper null model discounting the degree heterogeneity as well as other properties of the network \cite{bargigli2011random,squartini2011analytical}. 
Notably, the ERG framework can be directly used to generate networks with a community structure, by specifying the average numbers of links within and between each community. 
In this way the ensemble becomes equivalent to the classical {\em block-model}, in which each node is assigned to one of $B$ blocks (communities), 
and links are independently drawn between pairs of nodes with probabilities that are a function only of the block membership of the nodes \cite{fronczak2013exponential}. 
This means that eq. \eqref{eq:cm} becomes $p_{ij}=q_{b_ib_j}$, where $b_i$ denotes the block membership of node $i$, or $p_{ij}=(x_ix_jq_{b_ib_j})/(x_ix_jq_{b_ib_j}+1-q_{b_ib_j})$ 
for the degree-corrected block-model \cite{lancichenetti2008benchmark,karrer2011stochastic,peixoto2012entropy} in which also nodes' total degrees are constrained.

\begin{figure*}[p]
\includegraphics[width=\textwidth]{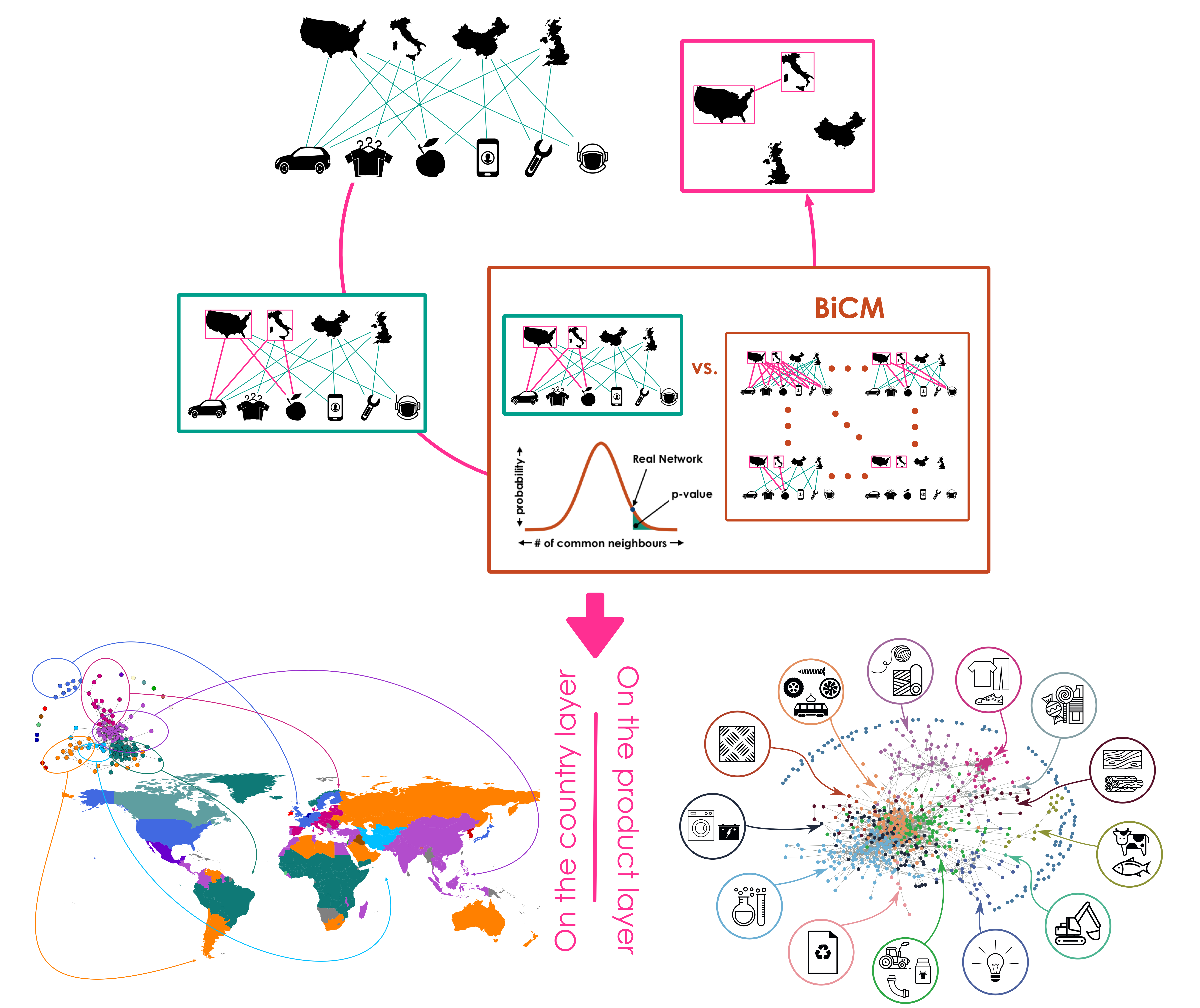}
\caption{One-mode projection of the network of countries and products they export (the bipartite representation of the WTW) and its statistical validation against a null hypothesis 
derived from the bipartite configuration model (BiCM). In the upper part of the figure, we illustrate the procedure applied on the countries set. 
First of all, the BiCM ensemble is derived constraining the degrees of both countries and products. The result is a connection probability between any country-product pair 
analogous to that of eq.~\eqref{eq:cm}. Then, taken two countries, the actual number of products they both export (cyan box) is compared with the expectation 
and probability distribution obtained from the BiCM (orange box). 
If this observed value is statistically significant (\ie, the null hypothesis that such value is explained by the degrees of both countries and products is rejected), 
a link connecting the two countries is established in the one-mode validated projection (magenta box). 
On the bottom left, we report results of the projection performed on the country set of the real WTW. The validation procedure helps to identify communities of countries with a similar industrial system. 
On the bottom right, we instead show that when the projection is performed on the product set, the statistical test highlights products requiring similar technological capabilities. 
Figure readapted from \cite{saracco2017inferring}. All icons are from the Noun Project and under the CC licence available at \url{http://thenounproject.com}.
\footnote{``China'' by Alexander Skowalsky; ``Italy'' by Mehmet I K Berker; ``United Kingdom'' by Luke Peek; ``United States'' by DeeAnn Gray; ``Apple'' by Kimberly Chandler; ``Car'' by Olivier Guin; 
``Clothes'' by Daniel Hanly; ``Smartphone'' by Gregor Cresnar; ``Screwdriver'' by FR; ``Astronaut Helmet'' by Yoshe; 
``Cow'' by Nook Fulloption; ``Fish'' by Iconic; ``Excavator'' by Kokota; ``Light bulb'' by Hopkins; ``Milk'' by Artem Kovyazin; 
``Curved Pipe'' by Oliviu Stoian; ``Tractor'' by Iconic; ``Recycle'' by Agus Purwanto; ``Experiment'' by Made; ``Accumulator'' by Aleksandr Vector; ``Washing Machine'' by Tomas Knopp; ``Metal'' by Leif Michelsen; 
``Screw'' by Creaticca Creative Agency; ``Tram'' by Gleb Khorunzhiy; ``Turbine'' by Luigi Di Capua; ``Tire'' by Rediffusion; ``Ball Of Yarn'' by Denis Sazhin; ``Fabric'' by Oliviu Stoian; 
``Shoe'' by Giuditta Valentina Gentile; ``Clothing'' by Marvdrock; ``Candies'' by Creative Mania; ``Wood Plank'' by Cono Studio Milano; ``Wood Logs'' by Alice Noir.}}
\label{fig:validation}
\end{figure*}

\paragraph*{\bf \em Bipartite networks and one-mode projections} --- Bipartite networks are a particular class of networks whose nodes can be divided into two disjoint sets, 
such that links exists only between nodes belonging to different sets \cite{holme2003network}. Typical examples of these systems include affiliation networks, 
where individuals are connected with the groups they are member of, and ownership networks where individuals are connected with the items they collected. 
The {\em Bipartite Configuration Model} (BiCM)~\cite{saracco2015randomizing} extends the BCM to this class of networks. 
This method has been used, for instance, to study the network of countries and products they export (\ie, the bipartite representation of the WTW) 
\cite{tacchella2012new,caldarelli2012network} and detect temporal variations related to the occurrence of global financial crises \cite{saracco2015detecting}. 
More recently, the BiCM has been applied to show that the degree sequence of interacting species in mutualistic ecological networks 
already induces a certain amount of {\em nestedness} of the interactions \cite{payrato2017breaking}.

In order to directly show the relation structure among one of the two sets of nodes, bipartite networks can be compressed into one-mode projection, 
namely a network containing only the nodes of the considered set---connected according to how many common neighbours they have in the other set \cite{zhou2007bipartite}. 
The problem of building a statistically validated projection of a bipartite network is similar in spirit to that of extracting the backbone from standard weighted networks 
\cite{tumminello2005tool,serrano2009extracting,slater2009two,radicchi2011information}, and has been typically addressed by determining which links are significant using a threshold---either 
unconditional or depending on the degree of nodes in the projected set \cite{goldberg2003assessing,latapy2008basic,tumminello2011statistically,tumminello2012identification,neal2013identifying}. 
However, differently from weighted networks, one-mode projections should be assessed against null models constraining the local information of both sets of the original bipartite network. 
Unfortunately, these models are much more difficult to derive. {\em Degree sequence models} used in the social sciences are based on computational link rewiring methods 
\cite{zweig2011systematic,horvat2013fixed} (which in this case are even more impractical and biased \cite{gionis2007assessing} than for standard networks) 
or require multiple observations of the empirical network \cite{neal2014backbone}. 
The null model for bipartite network projections derived from the maximum entropy principle is instead obtained by one-mode projecting the BiCM, 
that is, by computing the expected distribution for the number of common neighbours between nodes on the same layer \cite{gualdi2016statistically,saracco2017inferring} (Fig.~\ref{fig:validation}). 
Null models of these sort have been used, for instance, to analyse the one-mode projection of the bipartite WTW, allowing to detect modules of countries with similar industrial systems 
and a hierarchical structure of products \cite{saracco2017inferring}, as well as traces of specialisations emerging from the baseline diversification strategy of countries \cite{straka2017grand}. 
Or to study the patterns of assets ownership by financial institutions, identify significant portfolio overlaps bearing the highest riskiness for fire sales liquidation, 
and forecast market crashes and bubbles \cite{gualdi2016statistically}. More recently, a null model obtained by pairwise projecting multiple bipartite networks has been successfully applied 
to identify significant innovation patterns involving the interplay of scientific, technological and economic activities \cite{pugliese2017unfolding}.

\subsection*{Network reconstruction}

Many dynamical processes of critical importance, from the spread of infectious diseases to the diffusion of information and the propagation of financial losses, 
are highly sensitive to the topology of the underlying network of interactions \cite{pastorsatorras2015epidemic}. However in many situations the structure of the network is at least partially unknown. 
A classical example is that of financial networks: financial institutions publicly disclose their aggregate exposures in their balance sheets, 
whereas, individual exposures (\emph{who} is lending to whom and \emph{how much}) remain confidential \cite{wells2004financial,upper2011simulation,anand2017missing}. 
Another example is that of social networks, for which only aggregate information is released due to privacy issues, 
while their large scales deny the possibility of exhaustive crawling \cite{kossinets2006effects,lynch2008how}. For natural and biological networks, 
collecting all kinds of interactions is highly demanding because of technological limitations or high experimental costs \cite{amaral2008truer,guimera2009missing}. 
Thus reconstructing the network structure when only limited information is available is relevant across several domains, and represents one of the major challenges for complexity science. 

When the task is to predict individual missing connections in partly known networks one talks about {\em link prediction} \cite{lu2011link}. Here we instead discuss 
the fundamentally different task of reconstructing a whole network from partial information on the system, aggregated at the mesoscopic and macroscopic level \cite{squartini2018reconstruction}. 
The key to success is of course to make optimal use of what is known on the system, but also to make the most unbiased guess about what is not known. 
This is naturally achieved through the maximum entropy principle: the probability distribution which best represents the current state of knowledge on the network 
is the one with the largest uncertainty but satisfying the constraints corresponding to the available information. 
Note that the ERG approach has the additional advantage of not defining a single reconstructed network instance but an ensemble of plausible configurations with related probabilities. 
In this way, it can handle spurious or fluctuating data, and obtain robust confidence intervals for the outcomes of a given dynamical process on the (unknown) network.

Different kinds of local constraints lead to substantially diverse outcomes for the reconstruction process, though. 
Indeed, for a variety of networks of different nature (\eg, economic, financial, social, ecological networks), 
constraining the degrees as in the BCM typically returns a satisfactory reconstruction of the binary network features, whereas, 
constraining the strengths as in the WCM leads almost always to a very bad weighted reconstruction \cite{squartini2011analytical,mastrandrea2014enhanced}. 
The latter result is due to the entropy maximisation procedure being unbiased by not assuming any dependency between the strength of a node 
and the number of connections that node can establish. Hence, out of the many possible ways to redistribute the strength of each node over all possible links, 
the method chooses the most even one, so that the probability of assigning zero weight to a link is extremely small: 
the reconstructed network becomes almost fully connected---whatever the link density of the original network. 
This shows that degrees and strengths do carry different kinds of information, and constrain the network in a fundamentally different way. 
In order to reconstruct sparse weighted networks, both of them are required \cite{serrano2006correrlations,bhattacharya2008international}, as in the ECM \cite{mastrandrea2014enhanced}. 
This is quantitatively revealed using information-theoretic criteria (see \cite{mastrandrea2014enhanced} and Box 2).

\begin{mdframed}[backgroundcolor=ly]
\subsection*{Box 2: comparing models from different constraints}
Various likelihood-based statistical criteria exist to compare competing models resulting from different choices of the constraints, 
which are thus useful to assess the informativeness of different network properties. Consider two models $\mathcal{M}_1$ and $\mathcal{M}_2$. 
When these models are {\em nested}---meaning that $\mathcal{M}_2$ contains extra parameters with respect to $\mathcal{M}_1$, or equivalently that $\mathcal{M}_1$ is a special case of $\mathcal{M}_2$, 
comparison can be made through the {\em Likelihood Ratio Test} \cite{neyman1933problem}: 
if $\mbox{LRT}_{\mathcal{M}_1/\mathcal{M}_2}=-2[\mathcal{L}_{\mathcal{M}_1}(\vc{\theta}_1^*)-\mathcal{L}_{\mathcal{M}_2}(\vc{\theta}_2^*)]$ 
is smaller than a given significance level, $\mathcal{M}_2$ should be rejected even though its likelihood is by definition higher than that of $\mathcal{M}_1$, 
because $\mathcal{M}_2$ over-fits the data via redundant parameters that have inter-correlations and thus provide spurious information on the system \cite{burnham2002model}. 
When $\mathcal{M}_1$ and $\mathcal{M}_2$ are not nested, the {\em Akaike Information Criterion} \cite{akaike1974new} ranks them in increasing order of 
$\mbox{AIC}_{\mathcal{M}_r}=2[K_{\mathcal{M}_r}-\mathcal{L}_{\mathcal{M}_r}(\vc{\theta}_r^*)]$, where $K_{\mathcal{M}_r}$ is the number of parameters of model $\mathcal{M}_r$. 
For $R$ competing models, {\em Akaike weights} $\omega_{\mathcal{M}_r}=e^{-(\mbox{\tiny AIC}_{\mathcal{M}_r}/2)}/\sum_{r=1}^R e^{-(\mbox{\tiny AIC}_{\mathcal{M}_r}/2)}$ 
quantify the probability that each model is the most appropriate to describe the data \cite{wagenmakers2004aic}.
The {\em Bayesian Information Criterion} \cite{burnham2004multimodel} 
is similar to AIC, but accounts for the number $n$ of empirical observations as $\mbox{BIC}_{\mathcal{M}_r}=K_{\mathcal{M}_r}\ln n-2\mathcal{L}_{\mathcal{M}_r}(\vc{\theta}_r^*)$. 
\emph{Bayesian weights} can be also defined, analogously to Akaike weights. 
BIC is believed to be more restrictive than AIC \cite{burnham2002model}, yet which criteria performs best and under which conditions is still debated. 
Other model-selection methods have also been proposed, such as {\em multimodel inference} where a sort of average over different models is performed \cite{burnham2002model}.
\medskip
\end{mdframed}

\paragraph*{\bf \em The fitness ansatz} --- Unfortunately in many situations, typically in financial networks, also the node degrees are unknown. 
A possible solution here comes from the observation that, in many real networks, the connection probabilities between nodes can be expressed in terms of {\em fitnesses}, 
which are node ``masses'' typical of the system under analysis, so that node degrees can be mapped to their fitness values 
\cite{caldarelli2002scalefree,boguna2003class,garlaschelli2004fitness,garlaschelli2005scalefree,demasi2006fitness}. 
The strengths themselves work well as fitnesses in many cases \cite{barrat2004architecture}. The {\em fitness ansatz} then assumes that the strength of a node 
is a monotonic function of the Lagrange multiplier of the BCM controlling the degree of that node. 
Assuming the simplest linear dependency $\vc{s}^*\propto\vc{x}$ (but other functional forms are in principle allowed), 
eq. \eqref{eq:cm} becomes \cite{musmeci2013bootstrapping,cimini2015estimating,cimini2015systemic}
\begin{equation}
p_{ij}=\frac{zs_i^*s_j^*}{1+zs_i^*s_j^*}.\label{eq:fcm}
\end{equation}
The proportionality constant $z$ can be easily found using maximum likelihood arguments, together with a bootstrapping approach to assess the network density \cite{squartini2017network}. 
Node degrees estimated from such a fitness ansatz are then used to feed the ECM and obtain the ensemble of reconstructed networks \cite{cimini2015estimating}. 
Heuristic techniques can be alternatively employed to reduce the complexity of the method, by replacing the construction of the ECM ensemble 
with a {\em density-corrected gravity model} \cite{cimini2015systemic}. The methodology is also ready extendable to bipartite networks \cite{squartini2017enhanced}.

\begin{figure*}[p]
\centering
\includegraphics[width=0.5\textwidth]{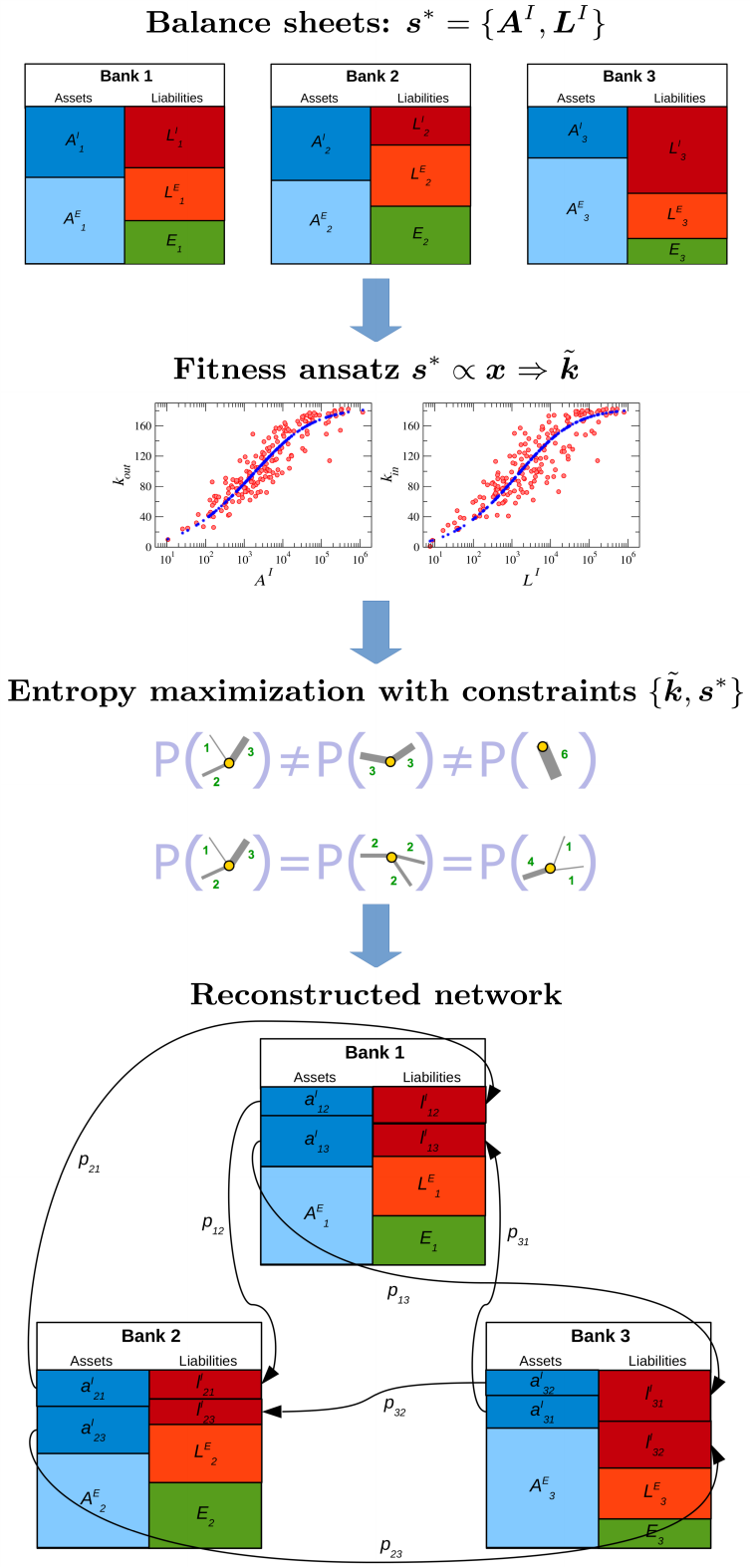}
\caption{Statistical reconstruction of an interbank network given by bilateral exchanges among banks. The method takes as input the exposure of each bank 
(\ie, its total interbank assets $A^I$ and liabilities $L^I$, obtained from publicly available balance sheets, which corresponds to the strength $\vc{s}^*$ of the node), 
and then performs two entropy maximisation steps. The first step estimates the (unknown) node degrees $\tilde{\vc{k}}$ and reconstructs the binary topology of the network 
using the fitness assumption that the number of connections and the total exposure of each bank scale together---resulting in a connection probability between nodes of the form of eq.~\eqref{eq:fcm}. 
This step requires as additional input an estimate of the density of the network, which can be obtained using a bootstrapping approach \cite{squartini2017network} 
relying on the hypothesis of statistical homogeneity (subsets of the network are representative of the whole system, regardless of the specific portion that is observed). 
The second step reconstructs the full weighted topology of the network, using either an ECM entropy maximisation constraining both degrees $\tilde{\vc{k}}$ estimated from the first step 
and empirical strengths $\vc{s}^*$ \cite{cimini2015estimating}, or a heuristic density-corrected gravity approach \cite{cimini2015systemic} to place weights on realised connections. 
The outcome of the reconstruction process is a probability distribution over the space of networks compatible with the constraints. 
Scatter plots adapted from \cite{cimini2015systemic}.}
\label{fig:reconstruction}
\end{figure*}

Note that despite having only the strengths as input, the described reconstruction method (Fig.~\ref{fig:reconstruction}) is different from the WCM 
by using these strengths not to directly reconstruct the network, but to estimate the degrees first, and only then to build the maximum entropy ensemble. 
In such a way it can generate sparse and non-trivial topological structures, and can faithfully be used to reconstruct complex networked systems \cite{anand2017missing,squartini2018reconstruction}.

\subsection*{Beyond local constraints}

ERG models are analytically tractable or not depending on whether a closed-form expression of the partition function $Z$ can be derived. 
As we have seen, this is indeed the case of local linear constraints, for which $Z$ factorises into link-specific terms. 
In some other cases, it may be possible to get approximate analytic solutions using a variety of techniques (mean-field theory, saddle-point approximation, diagrammatic perturbation theory, 
path integral representations). These situations have been vastly explored in the literature, and include 
the degree-correlated network model \cite{berg2002correlated}, the reciprocity model and the two-star model \cite{park2004solution,yin2016reciprocity}, the Strauss model of clustering \cite{park2005solution}, 
models of social collaboration \cite{fronczak2007phase}, models of community structure \cite{bianconi2008entropy}, hierarchical topologies \cite{bianconi2008entropies}, 
models with spatial embedding \cite{bianconi2009entropy} and rich-club features \cite{mondragon2014network}, 
and finally model constraining both the degree distribution and degree-degree correlations---which 
are known under the name of {\em Tailored Random Graphs} \cite{annibale2009tailored,roberts2011tailored,roberts2014entropies}. 
At last, when any analytic approach for computing $Z$ becomes intractable, the ensemble can be still populated using Monte Carlo simulations, either to explicitly sample the configuration space---taking care of 
avoiding sampling biases through the use of ergodic Markov chains fulfilling detailed balance \cite{artzy2005generating,coolen2009constrained,roberts2012unbiased}, 
or to derive approximate maximum likelihood estimators---taking care of avoiding degenerate regions of the phase space leading to frequent trapping in local minima 
\cite{strauss1990pseudolikelihood,vanduijn2009framework,snijders2010maximum,schweinberger2011instability,desmarais2012,chatterjee2013estimating,horvat2015reducing}. 
Such a variety of possible techniques is what makes ERG an extremely ductile and powerful framework for complex network modelling. 

\paragraph*{\bf \em Markov chain Monte Carlo} --- The usual Markov chain Monte Carlo (MCMC) method for ERG works as follows. Starting from a network $\mx{G}\in\Omega$, 
a new network $\mx{G}'\in\Omega$ is proposed by choosing two links at random and shuffling them as to preserve a given network property (such as the degree sequence of the network). 
The proposed network is accepted with the Metropolis-Hastings probability $Q_{\mx{G}\to\mx{G}'}=\min\{1,e^{H(\mx{G})-H(\mx{G}')}\}$ \cite{hasting1970montecarlo}, 
where $H$ is the Hamiltonian containing the Lagrange multipliers---the latter taking the role of inverse temperature(s) used for simulated annealing. 
The process is repeated from $\mx{G}'$ ($\mx{G}$) if the proposal is accepted (rejected). 
Since the moves fulfil ergodicity and detailed balance, for sufficiently long times the values of the constraints in the sampled networks 
are distributed according to what the canonical ensemble prescribes. However, despite this asymptotic guarantee, in practice this method often fails because the time to approximate 
the probability distribution grows exponentially with the system size. This happens whenever $P(\mx{G})$ possesses more than one local maximum, 
for instance when aiming at generating ensembles of networks with desired degree distribution, degree-degree correlations and clustering coefficient 
(within the so-called $dk$-series approach) \cite{mahadevan2006systematic,orsini2015quantifying}. Indeed rewiring methods biased by aiming at a given clustering coefficient 
display strong {\em hysteresis} phenomena: cluster cores of highly interconnected nodes do emerge during the process, but once formed they are very difficult to remove 
in realistic sampling time scales---leading to a break of ergodicity \cite{foster2010communities}. Multi-canonical sampling has been recently proposed to overcome this issue of phase transitions 
\cite{fischer2015sampling}. The idea is to explore the original canonical ensembles without being restricted to the most probable regions, which means sampling networks uniformly 
on a predefined range of constraints values. This is achieved using Metropolis-Hastings steps based on a microcanonical density of states estimated using the Wang-Landau algorithm \cite{wang2001efficient}.

\subsection*{Generalised network structures}

\begin{figure*}[t]
\centering
\includegraphics[width=0.5\textwidth]{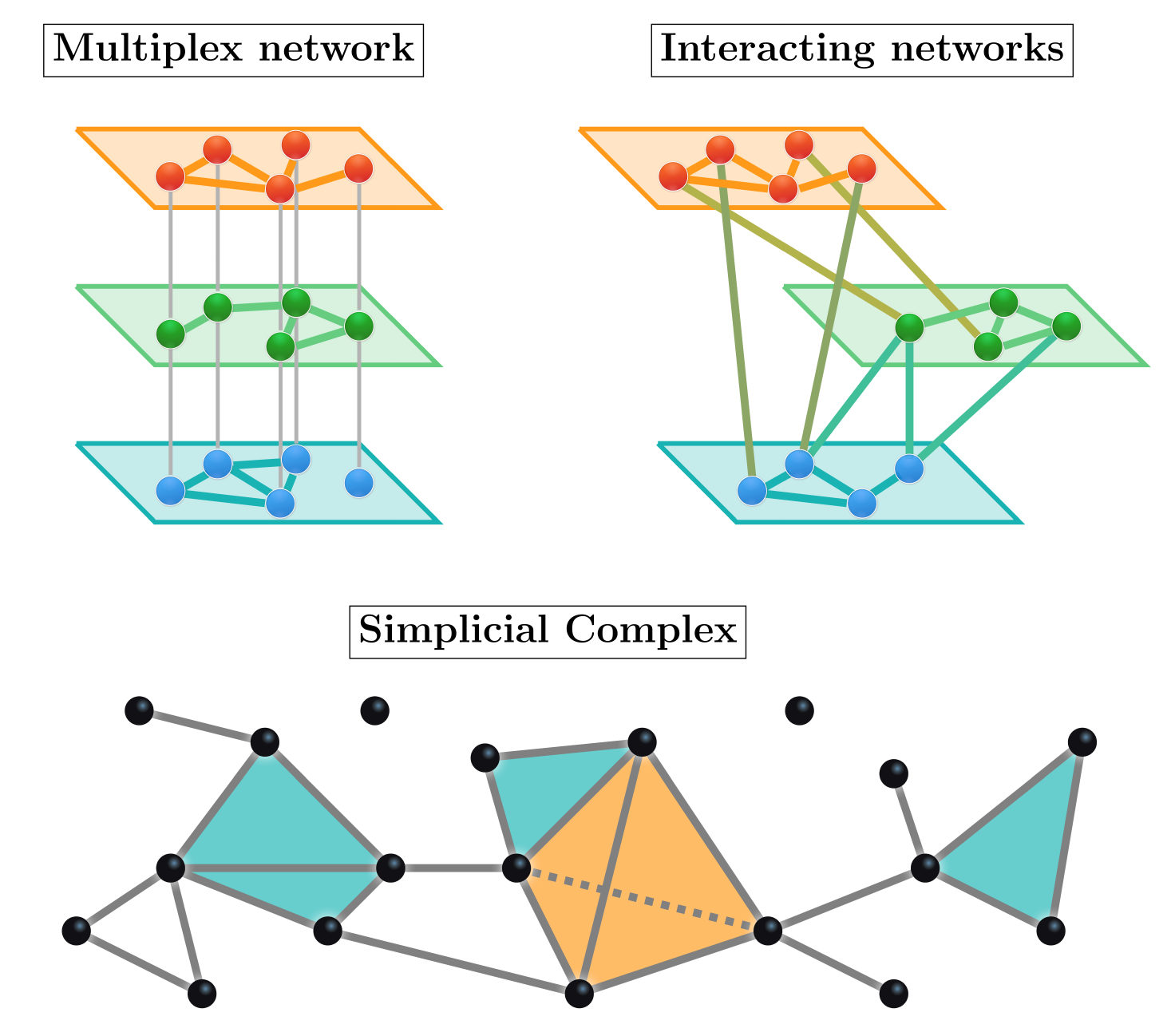}
\caption{Generalised network structures. A {\em multiplex} (or link-coloured) network consists of various layers with the same set of nodes and specific interaction patterns in each layer. 
An {\em interacting} (or multi-layer) network consists of various layers each with its own nodes, and of interactions existing within and between layers. 
A {\em simplicial complex} represents the different kind of interactions (simplices) between groups of $d$ nodes: 
isolated nodes ($d=0$), pairwise interactions ($d=1$, in grey), triangles ($d=2$, in cyan), tetrahedra ($d=3$, in yellow) and so on.}
\label{fig:generalised}
\end{figure*}

\paragraph*{\bf \em Networks of networks} --- Many complex systems are not just isolated networks, but are better represented by a ``network of networks'' 
\cite{kivela2014multilayer,boccaletti2014structure,dedomenico2016physics} (Fig.~\ref{fig:generalised}). 
The simplest and most studied situation is the so-called {\em multiplex}, where the same set of nodes interacts on several layers of networks. 
This is the case, for instance, of social networks where each individual has different kinds of social ties, 
or urban systems where locations can be connected by different means of transportation, 
or financial markets where institutions can exchange different kinds of financial instruments. 

Mathematically, a multiplex $\vec{\mx{G}}$ is a system of $N$ nodes and $M$ layers of interactions, each layer $\alpha=1,\dots,M$ consisting a network $\mx{G}^{\alpha}$. 
When modelling such a system, the zero-th order hypothesis is that the various layers are uncorrelated. 
In ERG terms, this means that the probability of the multiplex factorises into the probabilities of each network layer: 
$P(\vec{\mx{G}})=\prod_{\alpha=1}^M P_{\alpha}(\mx{G^{\alpha}})$ \cite{bianconi2013statistical,gemmetto2015multiplexity}. This happens whenever the constraints imposed on the multiplex 
are linear combination of constraints on individual network layers. Imposing local constraints separately for each network layer falls into this category. 
For instance, imposing the degrees in each layer leads to $M$ independent BCMs (one for each layer). The connection probability 
between any two nodes $i$ and $j$ in layer $\alpha$ reads $p_{ij}^{\alpha}=(x_i^{\alpha}x_j^{\alpha})/(1+x_i^{\alpha}x_j^{\alpha})$
where $\vc{x}^{\alpha}$ are layer-specific Lagrange multipliers, meaning that the existence of a link is independent on the presence of {\em any} other link in the multiplex. 
In this situation, the overlap of links between pairs of sparse network layers vanishes in the large $N$ limit. 

More realistic models of correlated multiplexes---in which the existence of a link in one layer is correlated with the existence of a link in another layer---can be 
generated by constraining the {\em multilink} structure of the system \cite{bianconi2013statistical}. 
A {\em multilink} $\vc{m}$ is an $M$-dimensional binary vector $\vc{m}$ indicating a given pattern of connections between a generic pair of nodes in the various layers. 
The {\it multidegree} $k(\vc{m})$ of a node in a given graph configuration is then the total number of other nodes with which the multilink $\vc{m}$ is realised. 
Constraining the multidegree sequence of the network (\ie, imposing $2^M$ constraints per node) leads to a probability for a multilink $\vc{m}$ between node $i$ and node $j$ given by 
$p_{ij}^{\vc{m}}=(x_i^{\vc{m}}x_j^{\vc{m}})/(\sum_{\vc{m}}x_i^{\vc{m}}x_j^{\vc{m}})$, which can be used to build systems made up of sparse layers with non-vanishing overlap 

Concerning weighted multiplexes, imposing the strength sequence (or degrees and strengths) on each layers is simple and again leads to uncorrelated layers \cite{menichetti2015weighted}. 
Models of correlated weighted multiplexes are instead obtained by constraining nodes multistrength, the weighted analogous of the multilink \cite{menichetti2014correlations}. 
An interesting and related case of study is provided by systems of aggregated multiplexes, \ie, simple networks given by the sum of the various layers of the multiplex. 
In this case, constraining the aggregated local structure leads to the traditional canonical ensemble. 
For instance, constraining the sum of the weights of connections incident to each node in each layer leads to the standard WCM. 
However, if we have information about the layered structure of the system---meaning that we know the number of layers of the original multiplex, 
the model can be built using a layer-degeneracy term for each link counting the number of ways its weight can be split across the layers. 
By doing so, one obtains a WCM($M$) model where the weight distribution is a negative binomial---the geometric distribution of the standard WCM 
being the special case $M=1$. Additionally, if we know that links in each layer possess different nature, a good modelling framework 
is equivalent to that of {\em multiedge} networks (ME) \cite{sagarra2013statistical,sagarra2014configuration}, where links belonging to different layers can be distinguished. 
To deal with this situation, it is necessary to introduce another degeneracy term counting all the network configurations giving rise to the same $P(\mx{G})$. 
The solution of the model is then obtained using a mixed ensemble with hard constraint for the total network weight and soft constraints for node strengths. 
This leads to a Poisson statistics (independent on $M$) for link weights and $\avg{w_{ij}}=My_iy_j\propto s^*_is^*_j$, 
a rather different situation than the outcome of either WCM or WCM($M$) \cite{sagarra2015role}. Now, it can be tricky to decide which statistics to use in a specific case. 
For instance, mobility or origin-destination networks, consisting of number of travels between locations (nodes) aggregated over observation periods (layers), 
are better modelled by ME networks---assuming that each trip is distinguishable. For the aggregated WTW the situation is less clear: 
while commodities are in principle distinguishable, trade transactions are much less so, and in fact neither the WCM, WCM($M$) and ME model can well reproduce it. 
A possible solution here is again to constrain both strengths and degrees simultaneously \cite{mastrandrea2014reconstructing}.

\paragraph*{\bf \em Simplicial complexes} --- Simplicial complexes are generalised network structures able to encode interactions occurring between more than two nodes, 
and allow to describe a large variety of complex interacting systems: collaboration networks where artefacts results from two or more actors working together, 
protein-interaction networks where complexes often consist of more than two proteins, economic systems of financial transactions often involving several parties, 
and social systems where groups of people are united by common motives or interest. Simplicial complexes can involve any number of nodes. 
For instance, simplices of dimension $d=0,1,2,3$ are, respectively, nodes, links, triangles, and tetrahedra, 
and in general a $d$-dimensional simplex is formed by a set of $d+1$ interacting nodes and includes all the simplices formed by subsets of $\delta+1$ nodes (with $\delta<d$), 
which are called the $\delta$-dimensional faces of the simplex. A simplicial complex is then a collection of simplices of different dimensions properly ``glued'' together (Fig.~\ref{fig:generalised}). 

Exponential random simplicial complexes (ERSC) have been recently introduced as a higher dimensional generalisations 
of ERG, and allow to generate random simplicial complexes where each simplex has its own independent probability of appearance, conditioned to the presence of simplex boundaries 
(which thus become additional constraints) \cite{zuev2015exponential}. 
In particular, the explicit calculation of the partition function is possible when considering random simplicial complexes formed exclusively by $d$-dimensional simplices 
\cite{owen2016generalized}. Indeed, by constraining the generalised degree (the number of $d$-dimensional simplices incident to a given $\delta$-dimensional face), 
the graph probability becomes a product of marginal probabilities for individual $d$-dimensional simplices. 
Alternatively, an appropriate Markov chain Monte Carlo sampling can be used to populate a microcanonical ensemble of simplicial complexes formed by simplices of any dimension \cite{young2017construction}.

\subsection*{Perspectives and Conclusion}

Complex networks have a twofold key difference with respect to the systems traditionally studied in equilibrium statistical physics.  
Firstly, their microscopic degrees of freedom are the interactions between the nodes making up the system, and not the states of the nodes themselves. 
Secondly, nodes are so heterogeneous, both in terms of intrinsic characteristics and connectivity features, that networks can hardly be assigned a typical scale. 
Maximum entropy models of networks based on local constraints are grounded on these two facts, by defining probability distributions on the network connections, 
and by not distinguishing nodes beyond their (heterogeneous) local features. 

As we reviewed here, these models have found an extremely wide range of practical applications. This is due to their analytic characterisation, 
and their versatility to encompass also higher-order network characteristics (\eg, assortativity, clustering, community structure) using stochastic sampling, 
as well as even more complex structures such as networks of networks and simplicial complexes. 
There are of course limitations of this approach, though. First of all, the constraints that can be imposed have to be static topological properties of the network. 
Only recently, cases of dynamical constraints have been considered using either the principle of Maximum Caliber---which is to dynamical pathways 
what the principle of maximum entropy is to equilibrium states \cite{presse2013principles,dixit2018perspective}, or definitions of the entropy functional alternative to Shannon's (see Box 3). 
Second, the possibility to consider in these models semi-local network properties heavily relies on numerical sampling, which becomes unfeasible or much biased 
for non-trivial patterns involving more than two or three nodes. These patterns can however be important in situations in which the network structure is determined by complex optimisation principles: 
sub-units of an electrical circuit, biochemical reactions in a cell, neuron firing patterns in the brain. 
Statistical physics of networks is bound to face the challenge of developing more complex network models for these kind of structures. 
Nevertheless, maximum entropy models based on local constraints do represent effective benchmarks to detect and validate them.

\begin{mdframed}[backgroundcolor=ly]
\subsection*{Box 3: Boltzmann, von Neumann, and Kolmogorov entropies}
For a microcanonical ensemble, the Boltzmann entropy is the logarithm of the number of network configurations belonging to the ensemble: $\Sigma=\ln\sum_{\mx{G}\in\Omega}\delta[\vc{c}(\mx{G}),\vc{c}^*]$. 
Similarly to the Shannon entropy for a canonical ensemble, the Boltzmann entropy can be used to quantify the complexity of the ensemble, \ie, 
to assess the role of different constraints in terms of information they carry on the network structure \cite{bianconi2008entropy,bianconi2008entropies}. 
Indeed, the more informative the constraints in shaping the network structure, the smaller the effective number of graphs with the imposed features, and so the lower the Boltzmann entropy 
of the corresponding ensemble. Using these arguments, it is possible to show that homogeneous degree distributions are the most likely outcome when the Boltzmann entropy of the microcanonical BCM ensemble is large, 
whereas, scale-free degree distributions naturally emerge for network ensembles with minimal Boltzmann entropy \cite{bianconi2009entropy}. 

The von Neumann entropy provides the amount of information encrypted in a quantum system composed by a mixture of pure quantum states 
\cite{braunstein2006laplacian,anand2009entropy,anand2011shannon,anand2014entropy}. 
It is given by $\Xi=-\mbox{Tr}(\mx{\rho}\ln\mx{\rho})$ where $\mx{\rho}$ is the density matrix of the system, and is thus equal to the Shannon entropy of the eigenvalue distribution of $\mx{\rho}$. 
For undirected binary networks, a formulation of this entropy which satisfies the sub-additivity property can be given in terms of the combinatorial graph Laplacian 
$\mx{L}=\mbox{diag}(\vc{k})-\mx{G}$, by defining $\mx{\rho}=e^{-\tau\mx{L}}/\mbox{Tr}(e^{-\tau\mx{L}})$ \cite{dedomenico2016spectral}. The resulting von Neuman entropy 
thus depends on the spectrum of the Laplacian, and can be seen as the result of constraining the properties of diffusion dynamics on the network \cite{delvenne2015diffusion,masuda2017random}.

Finally, the Kolmogorov entropy generalises the Shannon entropy by describing the rate at which a stochastic process generates information. 
Considering for simplicity a Markovian ergodic stochastic process, described by the matrix $\mx{\Phi}=\{\phi_{ij}\}$ of transition rates $\{i\to j\}$ 
and by the stationary asymptotic probability distribution $\{\pi_i\}$, the dynamical entropy is defined as $\kappa(\mx{\Phi})=-\sum_{ij}\pi_i\phi_{ij}\log \phi_{ij}$, 
that is the average of the Shannon entropies of rows of $\mx{\Phi}$---each weighted by the stationary distribution of the corresponding node \cite{demetrius2005robustness}. 
The Kolmogorov entropy turns out to be related on one hand to the capacity of the network to withstand random structural changes \cite{demetrius2005robustness}, 
and on the other hand to the Ricci curvature of the network \cite{lott2009Ricci}. This connection is particularly intriguing, 
as the Ricci curvature has been used to differentiate stages of cancer from gene co-expression networks \cite{sandhu2015graph}, 
as well as to give hallmarks of financial crashes from stock correlation networks \cite{sandhu2016Ricci}.
\medskip
\end{mdframed}

We remark that, from a practical point of view, the possibility of quantifying the relevance of a set of observed features and extracting meaningful information from huge streams of continuously-produced, 
high-dimensional noisy data is particularly relevant in the present era of Big Data. On one hand, the details and facets of information we are able to extract nowadays has reached levels never seen before, 
which means that we need more and more complex data structures and models in order to represent and comprehend them. 
On the other hand, the amount of information available calls for effective and scalable ways to let the signal emerge from the noise originated by the large variety of sources. 
In this effort, the theoretical framework of statistical physics stands as an essential instrument to make consistent inference from data, whatever the level of complexity we have to face.

\paragraph*{Acknowledgements} --- GCi, TS, FS and GCa acknowledge support from the EU projects CoeGSS (grant no. 676547), Openmaker (grant no. 687941), SoBigData (grant no. 654024) and DOLFINS (grant no. 640772). 
DG acknowledges support from the Dutch Econophysics Foundation (Stichting Econophysics, Leiden, the Netherlands). AG acknowledges support from the CNR PNR Project CRISISLAB funded by Italian Government.
GCa also acknowledges the Israeli-Italian project MAC2MIC financed by Italian MAECI.


%

\end{document}